\documentclass[aip,amsmath,amssymb,reprint]{revtex4-1}
\usepackage[dvipdfmx]{graphicx}
\usepackage{dcolumn}
\usepackage{bm}
\usepackage[utf8]{inputenc}
\usepackage[T1]{fontenc}
\usepackage{mathptmx}

\usepackage{color}

\begin{document}

\title{
  Screw dislocation that converts p-type GaN to n-type: Microscopic study on the Mg condensation and the leakage current in p-n diodes
}

\affiliation{ 
Graduate School of Engineering, Nagoya University, Nagoya 464-8601, Japan
}
\affiliation{ 
Institute of Materials and Systems for Sustainability, Nagoya University, Nagoya 464-8601, Japan
}
\affiliation{ 
Research Institute for Applied Mechanics, Kyushu University, Fukuoka 816-8580, Japan
}
\affiliation{
Toshiba Nanoanalysis Corporation, Yokohama 235-8522, Japan
}

\author{T. Nakano}
\email{nakano.takashi@h.nagoya-u.jp}
\affiliation{ 
Graduate School of Engineering, Nagoya University, Nagoya 464-8601, Japan
}

\author{Y. Harashima}
\altaffiliation[Corresponding author: ]{yosuke.harashima@imass.nagoya-u.ac.jp}
\affiliation{ 
Institute of Materials and Systems for Sustainability, Nagoya University, Nagoya 464-8601, Japan
}
\author{K. Chokawa}
\affiliation{ 
Institute of Materials and Systems for Sustainability, Nagoya University, Nagoya 464-8601, Japan
}
\author{K. Shiraishi}
\affiliation{ 
Graduate School of Engineering, Nagoya University, Nagoya 464-8601, Japan
}
\affiliation{ 
Institute of Materials and Systems for Sustainability, Nagoya University, Nagoya 464-8601, Japan
}
\author{A. Oshiyama}
\affiliation{ 
Institute of Materials and Systems for Sustainability, Nagoya University, Nagoya 464-8601, Japan
}
\author{Y. Kangawa}
\affiliation{ 
Research Institute for Applied Mechanics, Kyushu University, Fukuoka 816-8580, Japan
}
\author{S. Usami}
\affiliation{
Graduate School of Engineering, Nagoya University, Nagoya 464-8601, Japan
}
\author{N. Mayama}
\affiliation{
Toshiba Nanoanalysis Corporation, Yokohama 235-8522, Japan
}
\author{K. Toda}
\affiliation{
Toshiba Nanoanalysis Corporation, Yokohama 235-8522, Japan
}
\author{A. Tanaka}
\affiliation{ 
Institute of Materials and Systems for Sustainability, Nagoya University, Nagoya 464-8601, Japan
}
\author{Y. Honda}
\affiliation{ 
Graduate School of Engineering, Nagoya University, Nagoya 464-8601, Japan
}
\affiliation{ 
Institute of Materials and Systems for Sustainability, Nagoya University, Nagoya 464-8601, Japan
}
\author{H. Amano}
\affiliation{ 
Graduate School of Engineering, Nagoya University, Nagoya 464-8601, Japan
}
\affiliation{ 
Institute of Materials and Systems for Sustainability, Nagoya University, Nagoya 464-8601, Japan
}

\date{\today}

\begin{abstract}
Recent experiments suggest that Mg condensation at threading dislocations induce current leakage, leading to degradation of GaN-based power devices. 
To study this issue, we perform first-principles total-energy electronic-structure calculations for various Mg and dislocation complexes. 
We find that threading screw dislocations (TSDs) indeed attract Mg impurities, and that the electronic levels in the energy gap induced by the dislocations are elevated towards the conduction band as the Mg impurity approaches the dislocation line, indicating that the Mg-TSD complex is a donor. 
The formation of the Mg-TSD complex is unequivocally evidenced by our atom probe tomography in which Mg condensation and diffusion through [0001] screw dislocations is observed in p-n diodes. 
These findings provide a novel picture that the Mg being a p-type impurity in GaN diffuses toward the TSD and then locally forms an n-type region. 
The appearance of this region along the TSD results in local formation of the n-n junction and leads to an increase in the reverse leakage current.
\end{abstract}

\maketitle
The realization of next-generation power devices is a critical target for the sustainable development of society. 
A major role in electronics is currently played by Si-based power devices.~\cite{power_MOSFET,IGBT_1,IGBT_2,IGBT_3}
However, Si-based power devices are accompanied by large energy losses and there is a need to fabricate next generation devices that operate stably with low energy losses at high voltages. 
Gallium nitride (GaN) is one of the most promising materials since it has a larger band gap than Si and has a superior Baliga's figure of merit.
One can potentially reduce a large amount of the energy losses by replacing Si power devices with GaN 
and there are many reports already on the superior device performances.~\cite{GaN_MOSFET_1,GaN_MOSFET_2,GaN_vertical_MOSFET_1,GaN_vertical_MOSFET_2,GaN_vertical_MOSFET_3,GaN_vertical_MOSFET_4,GaN_vertical_MOSFET_5,GaN_vertical_MOSFET_6,GaN_vertical_MOSFET_7,Hu2015,Wang2018,Maeda2019}

In the power devices, p-n junctions are principal components which act as a barrier to prevent reverse bias leakage currents and then prevents the device from malfunctioning. 
The superior reverse bias characteristic is the important factor to guarantee the reliability and the performance of the devices.

In order to fabricate the GaN power devices, it is necessary to prepare free-standing GaN substrates and then to perform epitaxial growth. 
In practice, a high density of threading dislocations, which are extended from the substrate to the epitaxially grown GaN layers, have been observed.~\cite{edge_screw_mixed,Qian,Usami2018} 
For p-n diodes on a free-standing GaN substrate, the density of screw and mixed dislocations, penetrating the p-n junction, has been reported to be \(8\times10^{5}\ \rm cm^{-2}\) and \(6\times10^{6}\ \rm cm^{-2}\), respectively.~\cite{Usami2018} 
The presence of these dislocations is suspected as a primary source for the current leakage in the p-n diodes.~\cite{Usami2018} 

In addition to the intrinsic characteristics of the dislocations, 
the effect of an impurity-dislocation complex such as an Mg-dislocation complex in p-type GaN has been discussed in terms of its impact on the breakdown characteristics.~\cite{Akasaki_p-type} 
Previous experimental study\cite{Usami2019} has shown that Mg impurities are condensed along threading mixed dislocations (TMDs) in p-n diodes. 
The Mg configuration is most likely to be formed by a condensation of Mg at TMDs in a p-layer, followed by a subsequent diffusion of Mg through into a bottom n-layer along TMDs. 
The estimated Mg density was > \(10^{19}\ \rm cm^{-3}\) near the TMDs in the n-layer. 
It was revealed that the Mg diffusion along TMDs did not cause the current leakage. 
Usami \textit{et al}. have also reported the reverse leakage currents in vertical p-n diodes grown on GaN free-standing substrates\cite{Usami2018} and found that leakage spots correspond to the position of [0001] pure screw dislocations. 
The Mg condensation at the threading screw dislocations (TSDs), and its role on the leakage current has not been clarified so far.
Akiyama \textit{et al}.\cite{Akiyama_alloy} reported the atomic arrangement and the electronic structure of TSDs in \(\rm Al_{0.3}Ga_{0.7}N\) and \(\rm In_{0.2}Ga_{0.8}N\). 
However, there are no reports on the detailed atomic structures and the electronic properties of the Mg-dislocation complex in GaN. 
In order to reveal the origin of the reverse leakage current, it is necessary to understand Mg condensation mechanism and the electronic properties of the resulting Mg-TSD complex. 

In this Letter, we report results of first-principles calculations as well as atom probe tomography (APT) analysis that unveil microscopic origin of the condensation of Mg atoms towards the TSDs in GaN and of the observed reverse leakage current in p-n diodes. 
Our calculations unequivocally show that the formation of the Mg-TSD complex is energetically favorable and this complex induces electronic levels in the energy gap near the conduction-band bottom, hereby acting as a donor. 
We argue that this conversion of the Mg doped p-type GaN to the n-type occurring locally along the TSD is the reason for reverse leakage current in p-n diodes. 
The condensation of Mg atoms around the dislocation is clearly observed by APT. 

\begin{figure}
\includegraphics[width=\linewidth]{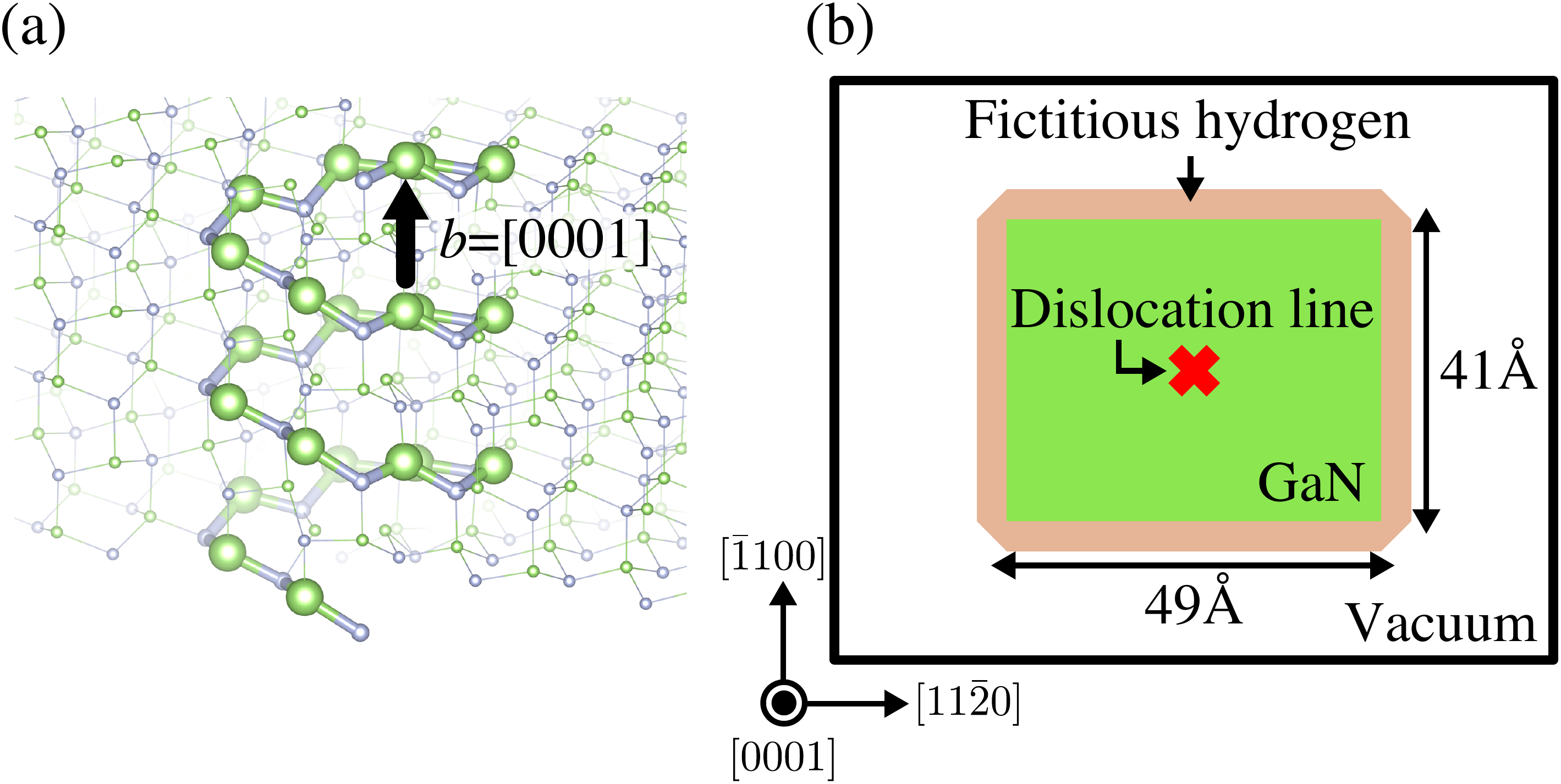}
\caption{\label{fig:model} (a) The model of the TSD. 
  The Burgers vectors is [0001] and dislocation line runs along [0001]. 
  Green and blue spheres are Ga and N atoms, respectively.
  The larger spheres highlight the dislocation core.
  (b) The schematic picture of the [0001] plane of the unit cell. 
  The green region denotes GaN and the surrounding is the fictitious hydrogen.
  The outermost region is the vacuum.
  A red cross denotes the position of the dislocation line.}
\end{figure}

All the calculations are performed based on the density functional theory (DFT).~\cite{Kohn_Sham_1,Kohn_Sham_2} 
We use the Vienna Ab initio Simulation Package (VASP).~\cite{VASP} 
The nuclei and core electrons are simulated by the pseudopotentials generated by the projector augmented-wave (PAW) method. 
The valence wave functions are expanded by the plane-wave basis set for which we find the cut-off energy of 450 eV suffices. 
The exchange-correlation energy is treated within the Generalized Gradient Approximation (GGA) with the Perdew-Burke-Ernzerhof functional.~\cite{GGA-PBE_1,GGA-PBE_2} 
In the electronic-structure calculations with the Heyd-Scuseria-Ernzerhof (HSE) hybrid functional\cite{HSE1,HSE2}, the amount of exact exchange is set to be 35 \% and the range-separation parameter is set to be 0.2 \AA$^{-1}$, leading to the band gap of 3.46 eV, which reproduces the experimental value for bulk GaN. 
Figure~\ref{fig:model}(a) shows an atomistic configuration of the TSD. 
The dislocation has the Burgers vector of [0001]. 
In order to model the TSD, the atoms are helically aligned along [0001] dislocation line at first.~\cite{screw_equ_1,screw_equ_2} 
A periodic boundary condition is then imposed on all axes of the system. 
The system contains 812 Ga and N atoms in total 
and the vacuum layer is added on lateral planes perpendicular to the [0001] direction~(see Fig. ~\ref{fig:model}(b)). 
The size except for the vacuum is 49\,\AA\; $\times$ 41\,\AA\; $\times$ 5.28 \AA.
The dangling bonds on the surfaces of GaN are terminated by `fictitious' hydrogen.~\cite{pseudoH} 
The integration over Brillouin zone is performed with 4 sampling points along the [0001] direction. 
Then the structural optimization is performed with the convergence criterion of \(5\times10^{-2}\) eV/\AA\; for all the atoms except for the fictitious H atoms which mimic the Mg-TSD complex in an otherwise infinite-size GaN crystal.

\begin{figure}[th]
\includegraphics[width=0.9\linewidth]{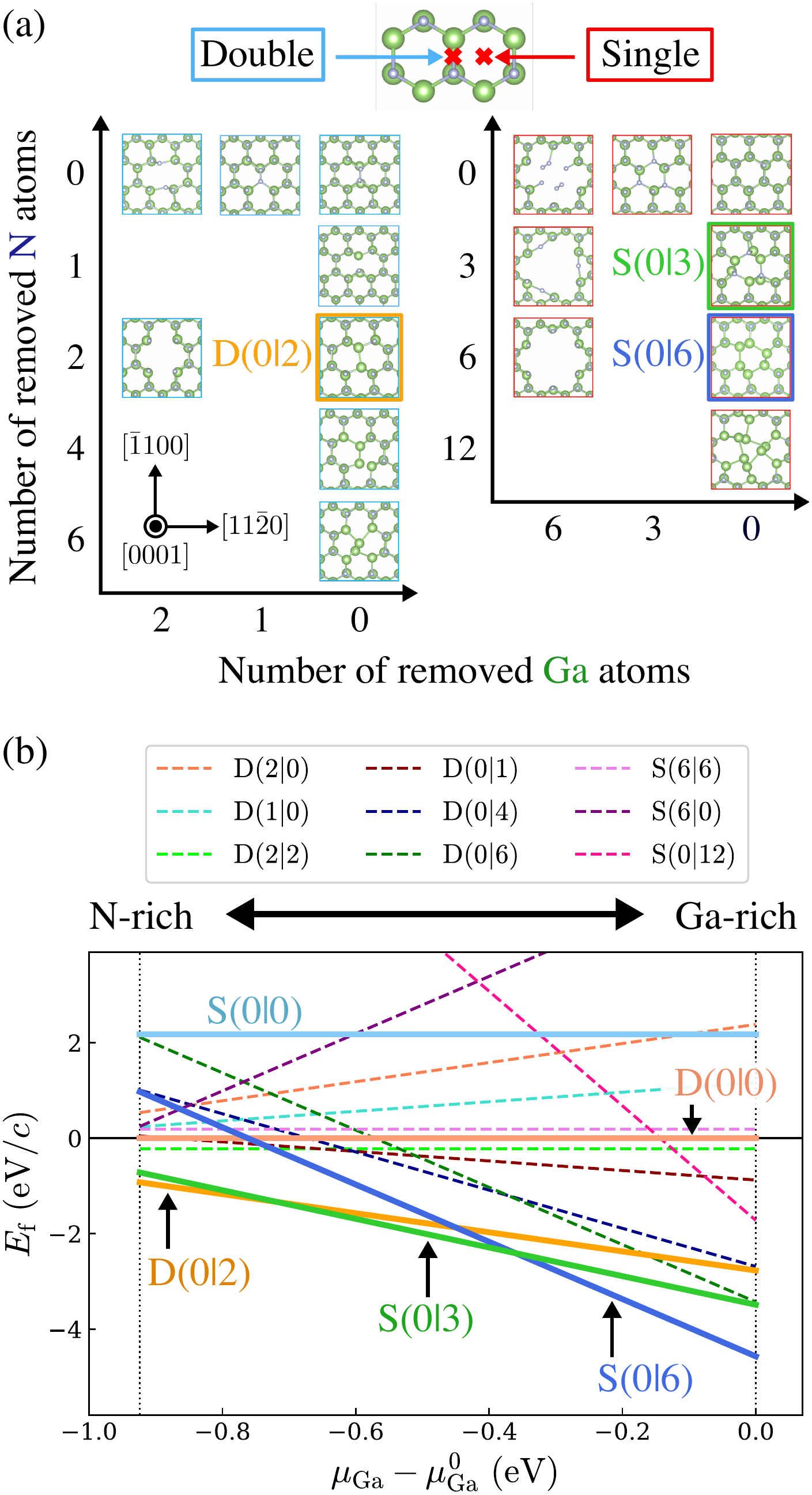}
\caption{\label{fig:core_type_mu-E} (a) Various relaxed core configurations of TSD seen from the [0001] direction. 
  A red cross in the top panel denotes the position of the dislocation line. 
  The left figure shows the double 6-atom ring core (labeled by ``D") in which the dislocation line is centered on a bond (basal plane projection). 
  The right shows the single 6-atoms ring core (labeled by ``S") in which the dislocation line is centered at a ring. 
  In both cases, different core stoichiometries are considered by removing atoms from the full core model. 
  Each core configuration is expressed as ``D($n$\,|\,$m$)" or ``S($n$\,|\,$m$)", where $n$ and $m$ denote the numbers of removed Ga and N atoms, respectively, from the corresponding filled core.
  (b) Formation energy per unit-cell length along the [0001] of various core configurations as a function of the Ga chemical potential. 
  The origin of the energy is chosen as the formation energy of the D(0\,|\,0).
  The values for the S(3\,|\,0) and the S(6\,|\,3) are too large to show in the figure, thus, omitted.}
\end{figure}

\begin{figure}
\includegraphics[width=\linewidth]{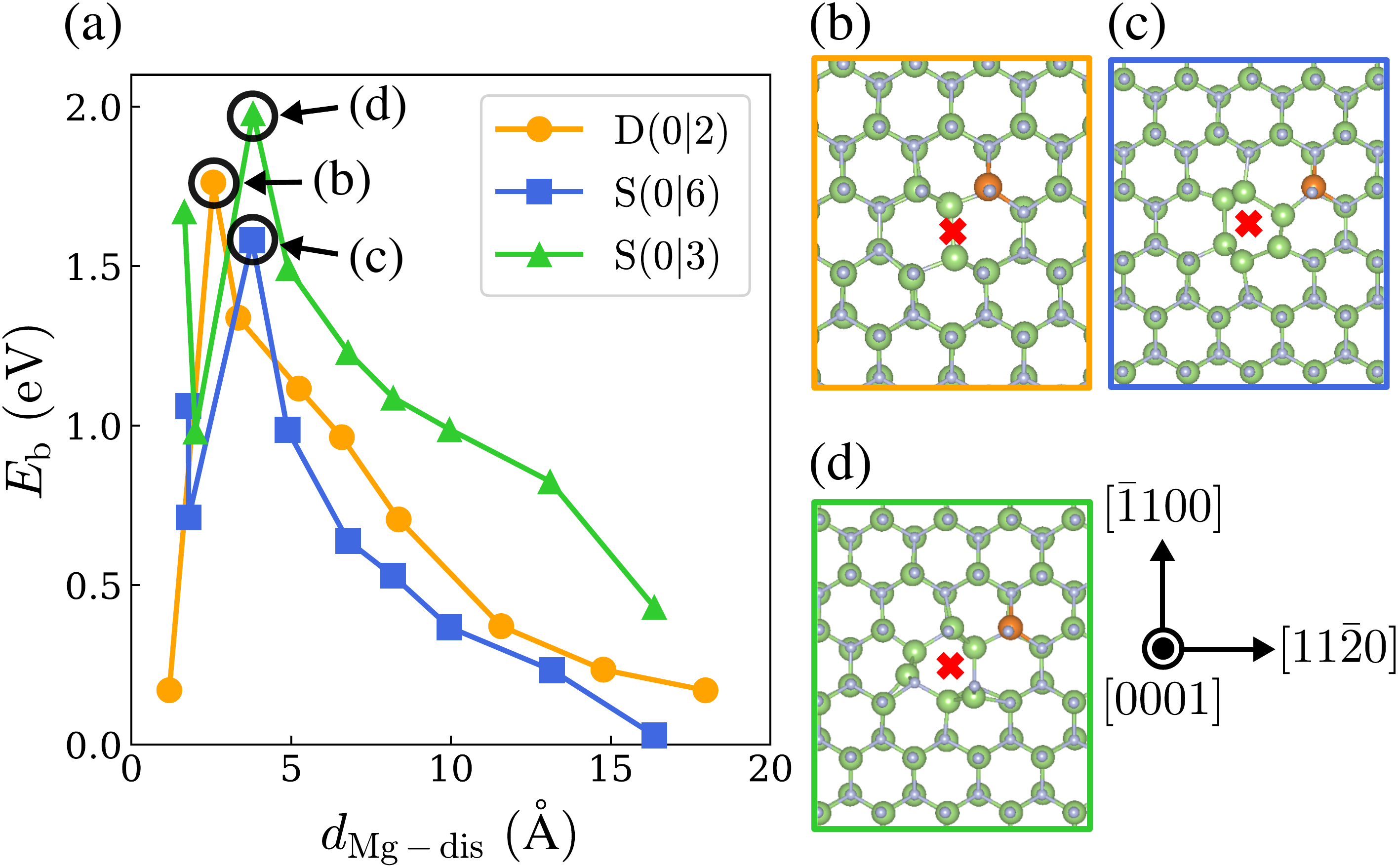}
\caption{\label{fig:binding_energy} 
  (a) The binding energy of the Mg-TSD complex of the D(0\,|\,2), the S(0\,|\,6), and the S(0\,|\,3) cores. 
  $d_{\text{Mg}-\text{dis}}$ is the distance of the Mg from the dislocation line. 
  (b--c) The most stable core structures of the Mg-TSD complexes of the D(0\,|\,2), the S(0\,|\,6), and the S(0\,|\,3) cores, respectively. 
  The position of the Mg atom is denoted by the orange sphere. 
  A red cross denotes the position of the dislocation line.}
\end{figure}

We set out the structural identification and the energetics of TSD cores with the Burgers vector \(\bm b =[0001]\) (for further details, see Supplementary Information). 
In addition to atomic reconstruction and relaxation near the core, we need to reveal how many Ga and N atoms are involved in the core region: from a fully filled core, partially filled cores and an open core. 
The previous first-principles calculations have examined only a part of possibilities.~\cite{Belabbas_ele,Northrup_Ga-filled,Groger_Ga-filled,core_energy_gain_1,Matsubara2013,Matsubara2014,Single_core} 
We perform systematic search for the most stable dislocation cores with possible 16 stoichiometric ratios of Ga and N as shown in Fig.~\ref{fig:core_type_mu-E}(a). 
The cores are expressed as, e.g., D($n$\,|\,$m$), where the first capital letter denote the position of the dislocation line, and $n$ and $m$ represent the numbers of Ga and N atoms, respectively, removed from the fully filled core.
The formation energy $E_{\text{f}}$ is defined as, 
\begin{equation}
  E_{\text{f}} = E-E^{\rm ref}-\mu_{\rm N}\Delta n_{\rm N}-\mu_{\rm Ga}\Delta n_{\rm Ga}.
\end{equation}
Here, \(E\) is the total energies of the supercell with the dislocation core and \(E^{\rm  ref}\) is the total energy of the reference model, for which we adopt the D(0\,|\,0) in the present study. 
\(\Delta n_{\rm N} (\Delta n_{\rm Ga} )\) is the difference in the numbers of N (Ga) atoms between the target and the reference models. 
The energetics depends on the chemical potential \(\mu\) of either Ga or N (Note that \(\mu_{\rm Ga}+\mu_{\rm N}=\epsilon_{\rm GaN}\) with \(\epsilon_{\rm GaN}\) being the energy of crystalline GaN per molecular unit). 
We have found three most stable cores of the TSD; D(0\,|\,2), S(0\,|\,6) and S(0\,|\,3) (Fig.~\ref{fig:core_type_mu-E}(b)). 
They are all Ga-rich cores with the formation energies about 1 eV per unit $c$ length (= 5.28 \AA) lower than other cores. 

We next clarify the nature of the interaction between an Mg atom and the stable TSDs obtained above. 
We replace a single Ga atom at various positions relative to the dislocation line with a single Mg atom. 
In our supercell model, the Mg concentration is \(1.0\times10^{20}\,\rm cm^{-3}\), which is comparable with the experimental values observed by APT analysis\cite{Usami2019} (\(\geqq 10^{19}\,\rm cm^{-3}\)). 
We define the binding energy \(E_{\rm b}\) between an Mg atom and a TSD as follows: 
\begin{equation}
E_{\rm b} = (E_{\rm dis}+E_{\rm bulk/Mg})-(E_{\rm dis/Mg}+E_{\rm bulk}),
\end{equation}
where \(E_{\rm dis/Mg}\), \(E_{\rm dis}\), \(E_{\rm bulk/Mg}\), \(E_{\rm bulk}\) are the total energies of the Mg-TSD complex, the dislocation without Mg, the Mg without the dislocation, and the system without the dislocation and Mg, respectively.
Figure~\ref{fig:binding_energy}(a) shows the binding energy as a function of the distance between the Mg atom and the dislocation line for the D(0\,|\,2), the S(0\,|\,6), and the S(0\,|\,3) cores. 
The binding energy increases as the Mg atom approaches the dislocation line for all the core structures. 
For all the 3 Ga-rich cores, D(0\,|\,2), S(0\,|\,6), and S(0\,|\,3), examined above, the most stable positions of the Mg atoms are next neighbor to the dislocation lines (see Figs. \ref{fig:binding_energy}(b)-\ref{fig:binding_energy}(d)). 
The Mg-TSD complex is lower in energy by 1.58 -- 1.76 eV per unit $c$ length than the isolated Mg and the dislocation.
These results unequivocally show that the TSDs attract Mg impurities and the resultant Mg-TSD complexes have shapes of rods with particular radii which are observable by the atom-probe experiment (see below).
\begin{figure}
\includegraphics[width=\linewidth]{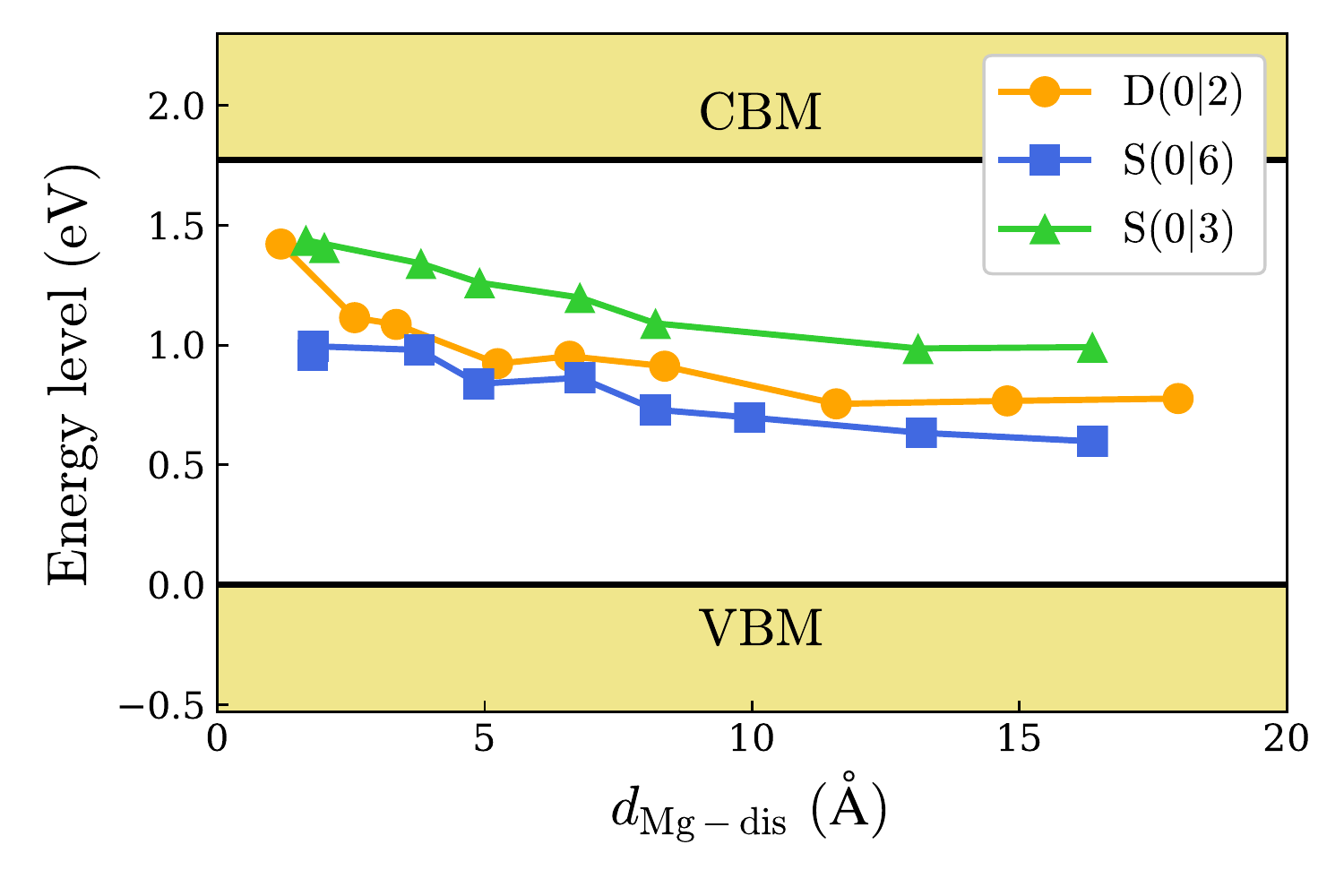}
\caption{\label{fig:highest_occupied_level} Distance dependence of the HOKS levels of the systems with the Mg-TSD complexes. 
  CBM and VBM denote conduction band minimum and valence band maximum, respectively. }
\end{figure}

We then reveal the electronic structures in the energy gap for the Mg-TSD complexes. 
We use Kohn-Sham (KS) single electron levels in DFT which approximate to proper electronic levels defined by using the total-energy difference between different charge states.~\cite{Baraff1980} 
Figure~\ref{fig:highest_occupied_level} shows the highest occupied KS (HOKS) levels of the Mg-TSD complexes for the D(0\,|\,2), the S(0\,|\,6), and the S(0\,|\,3) cores. 
In order to confirm the exact position of HOKS levels, we perform the electronic-structure calculations for the closest Mg-D(0\,|\,2) complex with the HSE functional.~\cite{HSE1,HSE2} 
We have obtained the HOKS level by using HSE functional located just below the conduction band minimum by an amount of 0.86 eV. 
The band gap estimated with GGA is about half of that with HSE, and the depth of this HOKS level with GGA is also half of that with HSE; that is, it is approximately scaled.
We found a clear trend that the HOKS levels are elevated toward the conduction band minimum as the Mg atom approaches the dislocation line, regardless of core structures. 
This result uncovers a fact unknown before that the Mg impurity, an acceptor when isolated, is coupled with the dislocation core and becomes a donor. 
This striking feature offers a new framework to consider physics and chemistry of defect-impurity complexes. 
To be specific to the present issue, the obtained results clearly indicate that Mg-doped p-GaN with TSDs induces n-type regions along the dislocation cores and thereby leads to the conduction collapse in p-n diodes. 
\begin{figure}
\includegraphics[width=\linewidth]{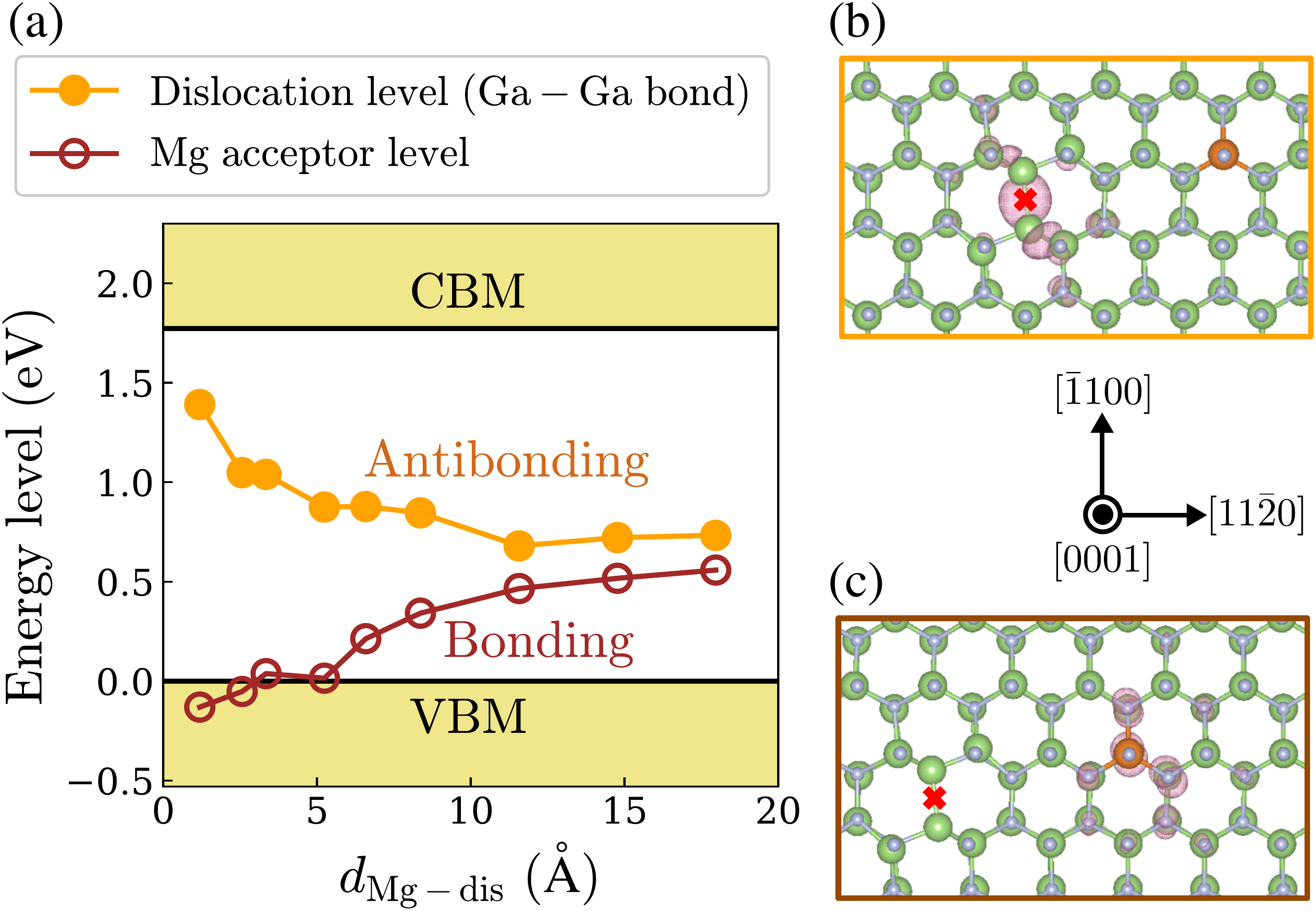}
\caption{\label{fig:hybridization} (a) Distance dependence of defect levels in the gap. 
  We plot the band centers of the dislocation states and the acceptor states.
  (b) The charge density of the state localized around a Ga-Ga bond crossing the dislocation line (sampled at \(\Gamma\) point, isosurfaces\,=\,0.005 electron/\AA\(^3\)). 
  (c) The charge density of the Mg acceptor state.}
\end{figure}

We are now in a position to discuss the physical reason for the elevation of the HOKS levels of the Mg-TSD complexes. 
We here focus on the Mg-D(0\,|\,2) complex but the physics is found to be the same for other complexes. 
By analyzing the KS levels in the energy gap, we have found that those levels in the gap are qualitatively classified into two groups: One is the levels originated from the core of the TSD which are found to be located in the mid gap, and the other is of course the levels coming from Mg impurities located near the valence-band top. 
The HOKS level has a character of the former. 
This is evidenced by our calculations shown in Fig.~\ref{fig:hybridization}(b) and (c) with the separation of 8.36 \AA\ between the Mg and the dislocation line. 
When the Mg impurity and the dislocation core becomes close, which is energetically favorable as stated above, the HOKS level shifts upward whereas the lower level does downward (Fig.~\ref{fig:hybridization}(a)). 
Our analyses of the KS orbitals show that the HOKS state has a character of the anti-bonding nature of the dislocation state and the Mg state, whereas the lower state has the bonding nature. 
The elevation of the HOKS level is therefore the upward shift of the anti-bonding state due to the enhanced hybridization.

\begin{figure}
\includegraphics[width=0.73\linewidth]{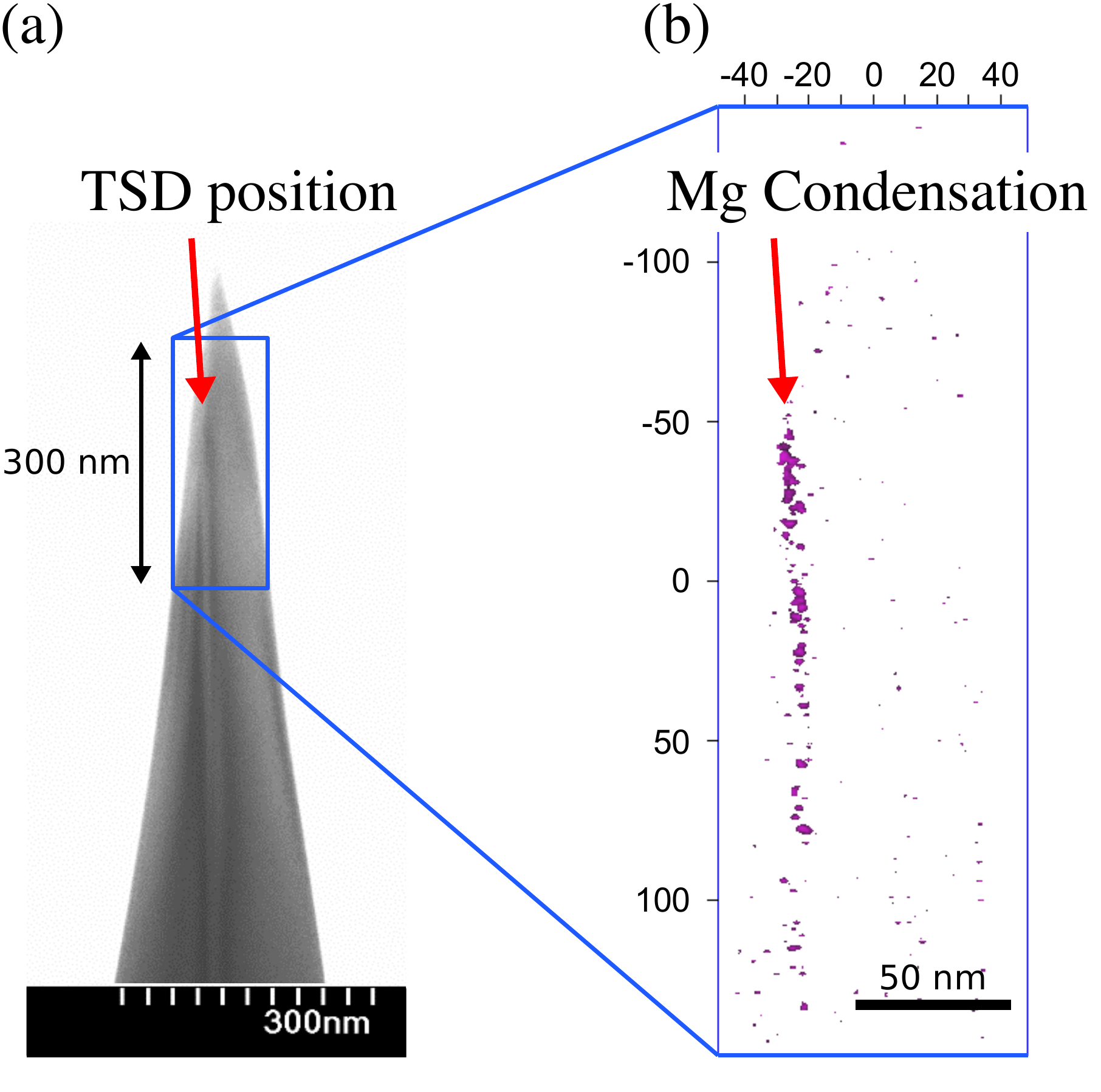}
\caption{\label{fig:Mg_diffustion} (a) The STEM image of the needle specimen of a TSD. 
  (b) The XZ plane projection view of the three-dimensional tomographic image of Mg 
  represented by the isoconcentration surface of 0.4 at. \% of Mg. }
\end{figure}

The theoretical findings described above are clearly corroborated by ATP experiments presented below. 
In order to observe the condensation of Mg atoms near the TSDs, we perform experimental analysis by using APT on Mg condensation in a sample with the TSD, which is discussed in Ref.~\onlinecite{Usami2018}. 
For details of the experiments, see Supplementary Information. 
Figure \ref{fig:Mg_diffustion}(a) shows the scanning transmission electron microscope (STEM) image of the needle specimen containing a TSD. 
Note here that the top of the needle specimen is not just below the pit due to the difficulty in capturing the center of the pit during the sample milling. 
The measured tomographic image of Mg, which shows isoconcentration surface with 0.4 at.\% of Mg, is shown in Fig. \ref{fig:Mg_diffustion}(b). 
The length of the specimen measured by APT is about 300 nm. 
The Mg tomographic image shows the Mg condensation along the dislocation. 
The observed diameter of the region with distributed Mg is less than 10 nm and the calculated concentration of Mg is over \(10^{19}\) \(\rm cm^{-3}\) around the dislocation. 
Therefore, it is considered that doped Mg impurities in p-type GaN diffuse into the n-type layer through TSDs. 
In Fig. \ref{fig:Mg_diffustion}(b), the Mg impurities tend to be condensed near the TSD in the n-type layer. 
Such highly concentrated Mg should come from the upper p-type layer. 
This result suggests that Mg impurities will also concentrate near the dislocation even in the p-type layer, which is consistent with our first-principles calculations.
Our experiment shows that the dislocation position where the Mg impurities are condensed coincides with the leakage spot. 
The APT analysis is consistent with our theoretical findings that the Mg-TSD complex is an origin of the leakage.

In summary, we have performed first-principles calculations that clarify stable core structures of screw dislocations in GaN and the formation of the complex of the screw dislocation and an Mg impurity atom for the first time. 
We have also found that these dislocation-Mg complexes act as donors, thus leading to the local conversion of p-type GaN to n-type. 
We argue that this conversion is the origin of the leakage current observed in the p-n diodes. 
Atom probe tomography has unequivocally corroborated this theoretical finding.  

\vspace{0.2in}
This work was supported by the MEXT ``Program for Research and Development of Next-Generation Semiconductors to Realize an Energy-Saving Society" under the contract number JPJ005357, and also by the MEXT Programs ``Social and Scientific Priority Issues Tackled by Post-K computer" and ``Promoting Researches on the Supercomputer Fugaku". 
The computation was partly conducted using the facilities of the Supercomputer Center, the Institute for Solid State Physics, the University of Tokyo.

The data that support the findings of this study are available from the corresponding author upon reasonable request.

\bibliography{reference}

\end{document}